\documentclass[twocolumn]{svjour3}          
\smartqed  
\usepackage{graphicx}

\usepackage{amsmath}
 \journalname{}
\begin{document}

\title{The impact of short term synaptic depression and stochastic vesicle dynamics on neuronal variability\thanks{This work supported by National Foundation of Science grant NSF-EMSW21-RTG0739261 and National Institute of Health grant NIH-1R01NS070865-01A1.}
}

\titlerunning{Short term depression and neuronal variability}        

\author{Steven Reich         \and
        Robert Rosenbaum 
}


\institute{Steven Reich \at
              Department of Mathematics, University of Pittsburgh, Pittsburgh PA, USA \\
           \and
           Robert Rosenbaum (corresponding author) \at
              Department of Mathematics and Center for the Neural Basis of Cognition, University of Pittsburgh, Pittsburgh PA, USA\\
              Tel.: +1412-624-8375\\
              Fax: +1412-624-83970\\
              \email{robertr@pitt.edu}  
}

\date{}

\maketitle

\begin{abstract}
Neuronal variability plays a central role in neural coding and impacts the dynamics of neuronal networks.  Unreliability of synaptic transmission is a major source of neural variability: synaptic neurotransmitter vesicles are released probabilistically in response to presynaptic action potentials and are recovered stochastically in time.  The dynamics of this process of vesicle release and recovery interacts with variability in the arrival times of presynaptic spikes to shape the variability of the postsynaptic response.   We use continuous time Markov chain methods to analyze a model of short term synaptic depression with stochastic vesicle dynamics coupled with three different models of presynaptic spiking: one model in which the timing of presynaptic action potentials are modeled as a Poisson process, one in which action potentials occur more regularly than a Poisson process and one in which action potentials occur more irregularly.  We use this analysis to investigate how variability in a presynaptic spike train is transformed by short term depression and stochastic vesicle dynamics to determine the variability of the postsynaptic response.  We find that regular presynaptic spiking increases the average rate at which vesicles are released, that the number of vesicles released over a time window is more variable for smaller time windows than larger time windows and that fast presynaptic spiking gives rise to Poisson-like variability of the postsynaptic response even when presynaptic spike times are non-Poisson.  Our results complement and extend previously reported theoretical results and provide possible explanations for some trends observed in recorded data.
\keywords{Short term depression \and Synaptic variability \and Fano factor}
\end{abstract}

\section{Introduction}
\label{intro}
Variability of neural activity plays an important role in population coding and network dynamics~\cite{Faisal08}.  Random fluctuations in the number of action potentials emitted by a population of neurons  affects the firing rate of downstream cells~\cite{Shadlen98,Shadlen98b}.  In addition, spike count variability over both short and long timescales can impact the reliability of a rate-coded signal~\cite{Dayan01}.  It is therefore important to understand how this variability is shaped by synaptic and neuronal dynamics.


Several studies examine the question of how intrinsic neuronal dynamics interact with variability in presynaptic spike timing to determine the statistics of a postsynaptic neuron's spiking response, 
but many of these studies do not account for dynamics and variability introduced at the synaptic level by short term synaptic depression and stochastic vesicle dynamics.  Synapses release neurotransmitter vesicles probabilistically in response to presynaptic spikes and recover released vesicles stochastically over a timescale of several hundred milliseconds~\cite{Zucker02,Fuhrmann02}.   The dynamics and variability introduced by short term depression and stochastic vesicle dynamics alter the response properties of a postsynaptic neuron~\cite{Vere1966,Abbott97,Chance98,Markram98,Goldman99,Senn01,Goldman02,Hanson02,Rocha04,Rocha05,Rothman09,Branco09,RosenbaumPLoS12} and therefore play an important role in information transfer~\cite{Zador98,Rocha02,Goldman04,Merkel10,Rotman11}, neural coding~\cite{Tsodyks97,Cook03,Abbott04,Grande05,Rocha08,Lindner09,Oswald12} and network dynamics~\cite{Tsodyks98,Galarreta98,Bressloff99,Wang99,Tsodyks00,Barbieri08}.  Understanding how variability in presynaptic spike times interact with short term depression and stochastic vesicle dynamics to determine the statistics of the postsynaptic response is therefore an important goal.



In this study, we use a model of short term synaptic depression with stochastic vesicle dynamics to examine how variability in a presynaptic input is transferred to variability in the synaptic response it produces.  We use the theory of continuous-time Markov chains to construct  exact analytical methods for calculating the the statistics of the postsynaptic response to three different presynaptic spiking models: one model with Poisson spike arrival times, one with more regular spike arrival times, and one with more irregular spike arrival times.
We find that depressing synapses shape the timescale over which neuronal variability occurs: the number of neurotransmitter vesicles released over a time interval is highly variable for shorter time windows, but less variable for longer time windows when variability is quantified using Fano factors.  Additionally, we find that when presynaptic inputs are highly irregular (Fano factor greater than 1), synaptic dynamics cause a reduction in Fano factor, consistent with previous studies~\cite{Goldman99,Goldman02,Rocha02,Rocha05}.  On the other hand, when presynaptic input is more regular (Fano factor less than 1), synaptic dynamics often cause an increase in Fano factor.  This observation suggests a mechanism through which irregular and Poisson-like variability can be sustained in spontaneously spiking neuronal networks~\cite{Tolhurst83,Softky93,Britten93,Buracas98,Mcadams99,Churchland10}, which complements previously proposed mechanisms~\cite{Vanreeswick96,Vanreeswijk98,Stevens98,Harsch00,Kumar12}.






\section{Methods}

We begin by introducing the synapse model used throughout this study.  We then proceed by analyzing the statistics of the synaptic response to three different input models.

\subsection{Synapse model}

A widely used model of depressing synapses~\cite{Tsodyks97,Abbott97,Tsodyks98,Markram98,Senn01} does not capture stochasticity in vesicle recovery and release.  As a result, this model underestimates the variability of the synaptic response~\cite{Rocha05,RosenbaumPLoS12}.  For this reason, we use a more detailed synapse model that takes stochastic recovery times and probabilistic release into account~\cite{Vere1966,Wang99,Fuhrmann02,RosenbaumPLoS12}.

We consider a presynaptic neuron with spike train $I(t)=\sum_j \delta(t-t_j)$ that makes $M$ functional contacts onto a postsynaptic cell.    Here, $t_j$ is the time of the $j$th presynaptic action potential.
Define $m(t)$ to be the number of contacts with a readily releasable neurotransmitter vesicle at time $t$ (so that $0\le m(t)\le M$).  
For simplicity, we assume that each contact can release at most one neurotransmitter vesicle in response to a presynaptic spike.
When a presynaptic spike arrives, each contact with a releasable vesicle releases its vesicle independently with probability ${p_r}$.  
After releasing a vesicle, a synaptic contact enters a refractory period during which it is unavailable to release a vesicle again until it recovers by replacing the released vesicle.  
The recovery time at a single contact is modeled as a Poisson process with rate $1/\tau_u$.  Equivalently, the duration of the refractory period is  exponentially distributed with mean $\tau_u$.

Define $w_j$ to be the number of  contacts that release a  vesicle in response to the presynaptic spike at time $t_j$ (so that $0\le w_j\le m(t_j^-)\le M$ where $m(t_j^-)=\lim_{t\to t_j^-}m(t)$).  The synaptic response is quantified by the marked point process
$$
x(t)=\sum_j w_j \delta(t-t_j).
$$
Since the signal observed by the postsynaptic cell is  determined by $x(t)$, we quantify synaptic response statistics in terms of the statistics of $x(t)$ in our analysis.  The process $x(t)$ can be convolved with a post-synaptic response kernel to obtain the conductance induced on the postsynaptic cell~\cite{RosenbaumPLoS12}.  The effects of this convolution on response statistics is well understood~\cite{Tetzlaff08}, so we do not consider it here.

This model can be described more precisely using the equation~\cite{RosenbaumPLoS12}
\begin{equation*}
dm(t)=-dN_x(t)+dN_u(t)
\end{equation*}
where $dN_u(t)=u(t)dt$ is the increment of an inhomogeneous Poisson process with instantaneous rate that depends on $m(t)$ through $\langle dN_u(t)\rangle\,|\, m(t)\rangle=dt(M-m(t))/\tau_u$ (here, $\langle\cdot\,|\,\cdot\rangle$ denotes conditional expectation), $N_x(t)=\int_0^t x(s)ds$ is the number of vesicles released up to time $t$, and each $w_j$ is a binomial random variable with mean ${p_r} m(t_j)$ and variance $m(t_j){p_r}(1-{p_r})$.

\subsection{Statistical measures of the presynaptic spike train and the synaptic response}

We focus on steady state statistics in this article, and therefore assume that the presynaptic spike trains are stationary and that the synapses have reached statistical equilibrium.
The intensity of a presynaptic spike train is quantified by the mean presynaptic firing rate, $$r_{\textrm{in}}=\langle I(t)\rangle=\langle N_I(T)/T\rangle$$ where $\langle \cdot \rangle$ denotes the expected value and $$N_I(T)=\int_0^T I(s)ds$$ represents the number of spikes in the time interval $[0,T]$.
Temporal correlations in the presynaptic spike times are quantified by the auto-covariance,
$$
R_{\textrm{in}}(\tau)={\textrm{cov}}(I(t),I(t+\tau)),
$$
and the variability in the presynaptic spike train is quantified by its Fano factor,
\begin{equation*}
F_{\textrm{in}}(T)=\frac{{\textrm{var}}(N_I(T))}{Tr_{\textrm{in}} }.
\end{equation*}
For much of this work, we will focus on Fano factors over large time windows which, through a slight abuse of notation, we denote by $F_{\textrm{in}}=\lim_{T\to\infty}F_{{\textrm{in}}}(T)$.
To compute Fano factors, we will often exploit their relationship to auto-covariance functions,
\begin{equation}\label{E:VarInt}
F_{\textrm{in}}(T)=\frac{1}{r_{\textrm{in}}}\int_{-T}^T R_{{\textrm{in}}}(\tau)(1-|\tau|/T)d\tau
\end{equation}
and
\begin{equation}\label{E:VarIntInf}
F_{\textrm{in}}=\frac{1}{r_{\textrm{in}}}\int_{-\infty}^\infty R_{{\textrm{in}}}(\tau)d\tau.
\end{equation}

The statistics of the synaptic response, $x(t)$, are defined analogously to the statistics of $I(t)$.
The steady state rate of vesicle release is defined as
$$
r_x=\langle x(t)\rangle=\langle N_x(T)/T\rangle
$$
where $N_x(T)=\int_0^T x(s)ds$ represents the number of vesicles released in  the time interval $[0,T]$.   Temporal correlations in the synaptic response are quantified by the auto-covariance, 
$$
R_x(\tau)={\textrm{cov}}(x(t),x(t+\tau))
$$
and response variability is quantified by the Fano factor of the number of vesicles released, 
$$
F_x(T)=\frac{{\textrm{var}}(N_x(T))}{Tr_x}.
$$
As above, we define $F_x=\lim_{T\to\infty}F_x(T)$ and note that
\begin{equation}\label{E:VarIntx}
F_x(T)=\frac{1}{r_{\textrm{in}}}\int_{-T}^T R_x(\tau)(1)d\tau \,\textrm{ and }\, F_x=\int_{-\infty}^\infty R_x(\tau)d\tau.
\end{equation}


\subsection{Model analysis with Poisson presynaptic inputs}

We first consider a homogeneous Poisson input, $I(t)$, with rate $r_{\textrm{in}}$.  The input auto-covariance for this model is given by $R_{{\textrm{in}}}(\tau)=r_{\textrm{in}} \delta(\tau)$ and the Fano factor is given by $F_{\textrm{in}}(T)=1$ for any $T> 0$.  
The mean rate of vesicle release for this model is given by
\begin{equation*}
r_x=\frac{M {p_r} r_{\textrm{in}}}{{p_r} r_{\textrm{in}} \tau _u+1}
\end{equation*}
which saturates to $M/\tau_u$ for large presynaptic rates, $r_{\textrm{in}}$.
A closed form approximations to the auto-covariance function of the response for this Poisson input model are derived in~\cite{RosenbaumPLoS12,Merkel10} (see also \cite{Rocha04}) and consist of a sum of a delta function and an exponential, 
\begin{equation}\label{E:RxPoisson}
R_x(\tau)=Dr_x\delta(\tau)-Er_xe^{-|\tau|/\tau_0},
\end{equation}
where the mass of the delta function is given by 
\begin{equation}\label{E:D}
D=\frac{2 {p_r} \left(r_{\textrm{in}} \tau _u+M-1\right)+2-{p_r}^2 r_{\textrm{in}} \tau _u}{(2-{p_r}) {p_r} r_{\textrm{in}} \tau _u+2}>0,
\end{equation}
the timescale of the exponential decay is given by
$$
\tau_0=\frac{\tau_u}{1+{p_r} r_{\textrm{in}}\tau_u},
$$
and the peak of the exponential is given by
\begin{equation}\label{E:E}
E=\frac{{p_r} r_{\textrm{in}} ((M-2) {p_r}+2) \tau _u+2 (M-1) {p_r}+2}{M (2-{p_r}) {p_r} r_{\textrm{in}} \tau _u+2 M}\; r_x.
\end{equation}
It can easily be checked that $E>0$ whenever $M\ge 1$, $0\le {p_r}\le1$, $r_{\textrm{in}}>0$, and $\tau_u>0$ so that the peak of the exponential in \eqref{E:RxPoisson} is negative.
For finite $T$, the Fano factor, $F_x(T)$, is  given by
 \begin{equation}\label{E:FxPoisson}
F_x(T)=D-2 E \tau _0-\left(e^{-\frac{T}{\tau _0}}-1\right)\frac{2 E \tau _0^2}{T}
\end{equation}
and, in the limit of large $T$, 
 \begin{equation}\label{E:FxInfPoisson}
F_x=D-2E\tau_0.
\end{equation}

To test the accuracy of these approximations, exact solutions can be found numerically using standard methods for the analysis of continuous-time Markov chains, as described for alternate input models below.  This analysis is a special case of the analysis for the ``regular'' input model described below that is achieved by taking $\theta=1$.  Alternatively, exact numerical results can be achieved by taking $r_s=r_f$ for the ``irregular'' input model.  In  figures showing results for the Poisson input model, we plot the closed form approximations described above along with exact numerical results obtained using the regular input model with $\theta=1$.

\subsection{Model analysis with irregular presynaptic inputs}  

Spike trains measured in vivo often exhibit  irregular spiking statistics indicated by Fano factors  larger than 1~\cite{Bair94,Dan96,Baddeley97,Churchland10}.  To describe the synaptic response to irregular inputs, we use a model of presynaptic spiking in which the instantaneous rate of the presynaptic spike train, $I(t)$, randomly switches between two values, $r_s$ and $r_f>r_s$, representing a slow spiking state and a fast spiking state.
The time spent in the slow  state before transitioning to the fast  state is exponentially distributed with mean $\tau_s$.  Likewise, the amount of time spent in the fast  state before switching to the slow state is exponentially distributed with mean $\tau_f$.  Transition times are independent from one another and from the spiking activity.  Between transitions, spikes occur as a Poisson process.

To find $r_{\textrm{in}}$, $R_{\textrm{in}}(\tau)$, and $F_{\textrm{in}}$, we represent this model as a doubly stochastic Poisson process.  Define $r(t)\in\{r_s,r_f\}$ to be the instantaneous firing rate at time $t$.  Then $r(t)$ is a continuous time Markov chain~\cite{Karlin1} on the state space $\Gamma=(r_s,r_f)$ with infinitesimal generator matrix
$$
A=\left[\begin{array}{cc} -1/\tau_s & 1/\tau_s\\ 1/\tau_f & -1/\tau_f\end{array}\right].
$$

Clearly, $r(t)$ spends a proportion $\tau_s/(\tau_s+\tau_f)$ of its time in the slow state (defined by $r(t)=r_s$) and a proportion $\tau_f/(\tau_s+\tau_f)$ of its time in the fast state (defined by $r(t)=r_f$).  This gives a steady-state mean firing rate of 
$$
r_\textrm{in}=\frac{r_s \tau_s+r_f\tau_f}{\tau_s+\tau_f}.
$$

At non-zero  lags ($\tau\ne 0$), the auto-covariance of a doubly stochastic Poisson process is the same as the auto-covariance of  $r(t)$~\cite{RosenbaumPLoS12}, which we can compute using techniques for analyzing continuous time Markov chains.  
For $\tau>0$, we have
\begin{align}
\langle r(t)r(t+\tau)\rangle&=r_s\Pr(r(t)=r_s)\langle r(t+\tau)\,|\,r(t)=r_s\rangle\notag\\ 
&\;\;+r_f\Pr(r(t)=r_f)\langle r(t+\tau)\,|\,r(t)=r_f\rangle \notag\\
&=\frac{r_s\tau_s}{\tau_s+\tau_f}\langle r(t+\tau)\,|\,r(t)=r_s\rangle\label{E:Err}\\
&\;\;+\frac{r_f\tau_f}{\tau_s+\tau_f}\langle r(t+\tau)\,|\,r(t)=r_f\rangle\notag
\end{align}
where $\langle \cdot |\cdot\rangle$ denotes  conditional expectation and 
\begin{align*}
&\langle r(t+\tau)\,|\,r(t)=r_s\rangle=r_s \Pr(r(t+\tau)=r_s\,|\,r(t)=r_s)\\
&\quad+r_f (1-\Pr(r(t+\tau)=r_s\,|\,r(t)=r_s)).
\end{align*}
The probability in this expression can be written in terms of an exponential of the generator matrix, $A$, and then calculated explicitly to obtain
\begin{align*}
\Pr(r(t+\tau)=r_s\,|\,r(t)=r_s)&=
\left[e^{A^T \tau}\left(\begin{array}{c}1\\0\end{array}\right)\right]_1\\
&=\frac{\tau _f e^{-\frac{\tau }{\tau _f}-\frac{\tau }{\tau _s}}+\tau _s}{\tau _f+\tau_s}
\end{align*}
where $[v]_k$ denotes the $k$th component of a vector, $v$.
An identical calculation can be performed to obtain an analogous expression for $\langle r(t+\tau)\,|\,r(t)=r_f\rangle$.  Combining these with Eq.~\eqref{E:Err} gives 
$$
\langle r(t)r(t+\tau)\rangle=\frac{\tau _f \tau _s \left(r_f-r_s\right)^2 }{\left(\tau _f+\tau _s\right)^2}e^{-\frac{\tau}{\tau_s}-\frac{\tau}{\tau_f}}+r_{\textrm{in}}^2.
$$
For positive $\tau$, we have $R_{\textrm{in}}(\tau)=\langle r(t)r(t+\tau)\rangle-r_{\textrm{in}}^2$.
As with all  stationary point processes $R_{\textrm{in}}(\tau)=R_{\textrm{in}}(-\tau)$ and $R_{\textrm{in}}(\tau)$ has a Dirac delta function with mass $r_{\textrm{in}}$ at the origin~\cite{Cox}.  Thus, the auto-covariance of $I(t)$ is given by
\begin{equation}\label{E:RinBurst}
R_{\textrm{in}}(\tau)=r_{\textrm{in}}\delta(\tau)+\frac{\tau _f \tau _s \left(r_f-r_s\right)^2 }{\left(\tau _f+\tau _s\right)^2}e^{-\frac{|\tau|}{\tau_s}-\frac{|\tau|}{\tau_f}}.
\end{equation}
For finite $T$, the Fano factor, $F_{\textrm{in}}(T)$, can  be computed using Eqs.~\eqref{E:VarInt} and \eqref{E:RinBurst}.  In the limit of large $T$, we can use Eqs.~\eqref{E:VarIntInf} and \eqref{E:RinBurst} to obtain a closed form expression,
\begin{equation}\label{E:FinBurst}
F_{\textrm{in}}=1+\frac{2 \tau _f^2 \tau _s^2 \left(r_f-r_s\right)^2}{\left(\tau _f+\tau _s\right)^2
   \left(r_f \tau _f+r_s \tau _s\right)}.
\end{equation}

Poisson spiking is recovered by setting $r_f=r_s$, $\tau_f=0$, or $\tau_s=0$.  
For any other parameter values (i.e., when $r_f\ne r_s$ and $\tau_f,\tau_s>0$),  it follows from Eq.~\eqref{E:FinBurst}  that $F_{\textrm{in}}(T)> 1$ for any $T$.  Therefore this input model, hereafter referred to as the ``irregular spiking'' model, represents spiking that is more irregular than a Poisson process.

The analysis in \cite{RosenbaumPLoS12} used to derive closed form expressions for the response statistics with Poisson inputs cannot easily be generalized to derive expressions with non-Poisson inputs like those considered here.  
Instead, we analyze the synaptic response for the irregular input model using techniques for analyzing continuous time Markov chains.
First note that the process $b(t)=(m(t),r(t))$ is a  continuous-time Markov chain on the discrete state space $\{0,1,\ldots,M\}\times \{r_s,r_f\}$.   
Here, $m(t)$ denotes the size of readily releasable pool and $r(t)$ represents the instantaneous presynaptic rate (which switches between $r_s$ and $r_f$).
We enumerate all $2(M+1)$ elements of this state space and denote the $j$th element of this enumeration as $\Gamma_j=(m_j,r_j)$ for $j=1,\ldots,2(M+1)$.


The infinitesimal generator, $B$, of $b(t)$ is a $2(M+1)\times 2(M+1)$ matrix with off-diagonal terms defined by the instantaneous transition rates,
\begin{equation}\label{E:B}
B_{j,k}=\lim_{h\to 0}\frac{1}{h}\Pr(b(t+h)=\Gamma_k\,|\,b(t)=\Gamma_j),\;\; j\ne k
\end{equation}
and with diagonal terms chosen so that the rows sum to zero: $B_{j,j}=-\sum_{k\ne j}B_{j,k}$~\cite{Karlin1}.

To fill the matrix $B$, we consider each type of transition that the process $b(t)$ undergoes.  
Vesicle recovery events occur at the instantaneous rate $(M-m(t))/\tau_u$ and increment the value of $m(t)$ by one vesicle.  Therefore 
$$
\lim_{h\to 0}\frac{1}{h}\Pr(b(t+h)=(m+1,r)\,|\,b(t)=(m,r))=\frac{M-m}{\tau_u}
$$
for $m\in\{0,\ldots,M-1\}$ and $r\in\{r_s,r_f\}$.  Vesicle release events occur at the instantaneous rate $r(t)$ and decrement the value of $m(t)$ by a random amount $k$ with a binomial distribution so that
\begin{align*}
\lim_{h\to 0}&\frac{1}{h}\Pr(b(t+h)=(m-k,r)\,|\,b(t)=(m,r))=\\
&r\, \frac{m!}{(m-k)!}{p_r}^k(1-{p_r})^{m-k}
\end{align*}
for $m\in\{1,\ldots,M\}$, $k\in \{0,\ldots,m\}$, and $r\in\{r_s,r_f\}$.
The value of $r(t)$ switches from $r_s$ to  $r_f$ with instantaneous rate $1/\tau_s$ so that
$$
\lim_{h\to 0}\frac{1}{h}\Pr(b(t+h)=(m,r_f)\,|\,b(t)=(m,r_s))=\frac{1}{\tau_s}
$$
and, similarly,
$$
\lim_{h\to 0}\frac{1}{h}\Pr(b(t+h)=(m,r_s)\,|\,b(t)=(m,r_f))=\frac{1}{\tau_f}.
$$
These four transition types account for all of the transitions that $b(t)$ undergoes.  They can be used to fill the off-diagonal terms of the matrix $B$.  The diagonal terms are then filled to make the rows sum to zero, as discussed above.  

Once a the infinitesimal generator matrix, $B$, is obtained, the probability distribution of $b(t)$ given an initial distribution $p(0)$ is given by
$$
p(t)=e^{t B^T}p(0).
$$
The stationary distribution, $p_0$, of $b(t)$ is given by the vector in the one-dimensional null space of $B$ with elements that sum to one~\cite{Karlin1}.  

The instantaneous rate of vesicle release, conditioned on the current state of $r(t)$ and $m(t)$, is given by
$$
\langle x(t)\,|\, r(t)=r,\,m(t)=m\rangle=r{p_r} m.
$$
Averaging over $r$ and $m$ in the steady state gives
$$
r_x=\sum_{j=1}^{2(M+1)}[p_0]_j r_j {p_r} m_j
$$
where $[\cdot]_j$ denotes the $j$th element.
The auto-covariance, $R_x(\tau)$, has a Dirac delta function at $\tau=0$.  We separate this delta function from the continuous part by writing
$
R_x(\tau)=A_x \delta(\tau)+R^+_x(\tau)
$
where $R^+_x(\tau)$ is a continuous function.  
The area of the delta function can be found by conditioning on the current state of $r(t)$ in the steady state to get
\begin{align}
A_x&=\lim_{t\to\infty}\langle x(t)^2dt\rangle\notag\\
&=\lim_{t\to \infty}\langle dN_x^2(t)/dt\rangle\notag\\
&=\sum_{j=1}^{2(M+1)}r_j[p_0]_j \lim_{k\to\infty} \langle w_k^2\,|\, m(t_k^-)=m_j\rangle\label{E:Ax1}
\end{align}
where $w_k$ is the number of vesicles released by the $k$th presynaptic spike.  Conditioned on the size, $m(t_k^-)$, of the readily releasable pool immediately before the presynaptic spike arrives, $w_k$ has a binomial distribution with second moment,
$$
\lim_{k\to\infty}\langle w_k^2 \,|\, m(t_k^-)=m_j\rangle=m_j{p_r} (1-{p_r} )+m_j^2{p_r}^2
$$
which can be substituted into Eq.~\eqref{E:Ax1} to calculate $A_x$.

All that remains is to calculate the continuous part, $R^+_x(\tau)$, of $R_x(\tau)$.
First note that, for $\tau>0$,
\begin{align}
&\lim_{t\to\infty}\langle x(t)x(t+\tau)\rangle\notag\\
&=\sum_{i,j=1}^{2(M+1)}[p_0]_i \Pr(b(t+\tau)=\Gamma_j\,|\,b(t)=\Gamma_i)\label{E:xxBurst}\\
&\times \langle dN_x(t)dN_x(t+\tau)\,|\,b(t)=\Gamma_i,b(t+\tau)=\Gamma_j\rangle/dt^2.\notag
\end{align}
The second term in Eq.~\eqref{E:xxBurst} can be computed as
$$
\Pr(b(t+\tau)=\Gamma_j\,|\,b(t)=\Gamma_i)=[e^{\tau B^T} {\textbf e}_{i}]_j=[e^{\tau B^T}]_{j,i}
$$
where ${\bf e}_i$ is the $2(M+1)\times 1$ vector whose $i$th element is 1 and all other elements are zero, which represents an initial distribution concentrated at $\Gamma_i$.  
The last term in Eq.~\eqref{E:xxBurst} is given by 
\begin{align*}
&\langle dN_x(t)dN_x(t+\tau)\,|\,b(t)=\Gamma_i,b(t+\tau)=\Gamma_j\rangle/dt^2=\\
&r_i r_j {p_r}^2 m_im_j.
\end{align*}
Finally,  $R_{x}^+(\tau)=\lim_{t\to\infty}\langle x(t)x(t+|\tau|)\rangle-r_x^2$ for $\tau\ne 0$ so that
\begin{align*}
&R_x(\tau)=\\
&A_x\delta(\tau)-r_x^2+\sum_{i,j=1}^{2(M+1)} [p_0]_i\left[e^{|\tau| B^T}\right]_{j,i}r_i r_j {p_r}^2m_im_j
\end{align*}
which can be computed efficiently using matrix multiplication.
The response Fano factor, $F_x$, can then be found by integrating $R_x(\tau)$ according to Eqs.~\eqref{E:VarInt} and \eqref{E:VarIntInf}.

%


\subsection{Model analysis with regular presynaptic inputs}

We now consider a spiking model that gives Fano factors smaller than 1 and therefore spike trains that are more regular than Poisson processes.  We achieve this by defining a renewal process with gamma-distributed interspike intervals (ISIs).  Such a process can be obtained by first generating a Poisson process, $\sum_k\delta(t-s_k)$ with rate $r=\theta\, r_{\textrm{in}}$ for some positive integer $\theta$, then keeping only every $\theta$th spike to build the spike train $I(t)$.
More precisely, the first spike of the gamma process is  obtained by choosing an integer, $k$, uniformly from the set $\{1,\ldots,\theta\}$ and defining 
defining $t_1=s_k$.  The remaining spikes are defined by $t_{j+1}=s_{j\theta +k}$ to obtain the stationary renewal process, $I(t)=\sum_j \delta(t-t_j)$~\cite{CoxRenewal}.

Clearly, this process has rate $r_{\textrm{in}}$ since the original Poisson process has rate $\theta r_{\textrm{in}}$ and  a proportion $1/\theta$ of these spikes appear in $I(t)$.  
The auto-covariance is given by~\cite{RosenbaumThesis}
\begin{equation}\label{E:RinGamma}
R_{\textrm{in}}(\tau)=r_{\textrm{in}} \delta(\tau)+r_{\textrm{in}}\left(\sum_{k=1}^\infty f_k(\tau)-r_{\textrm{in}}\right).
\end{equation}
where 
$$
f_k(t)=\frac{t^{k\theta-1}(\theta r_{\textrm{in}})^{k\theta} e^{-\theta r_{\textrm{in}} t}}{(k\theta-1)!}
$$
is the density of the waiting time between the first spike and the $(k+1)$st spike (i.e., the duration of $k$ consecutive ISIs).

For finite $T$, the Fano factor, $F_{\textrm{in}}(T)$, can  be computed using Eqs.~\eqref{E:VarInt} and \eqref{E:RinGamma}.  
In the limit of large $T$, we can use Eq.~\eqref{E:VarIntInf} or use the fact that, for renewal processes, $F_{\textrm{in}}={\textrm{var}}(\textrm{ISI})/\langle \textrm{ISI}\rangle^2$ where ${\textrm{var}}(\textrm{ISI})=1/(r_{\textrm{in}}^2\theta)$ is the variance and $\langle \textrm{ISI}\rangle=1/r_{\textrm{in}}$ is the mean of the gamma distributed ISIs~\cite{CoxRenewal}.  This gives
\begin{equation*}
F_{\textrm{in}}=\frac{1}{\theta}.
\end{equation*}
Poisson spiking is recovered by setting $\theta=1$.  When $\theta>1$,  we have that $F_{\textrm{in}}(T)< 1$ for any $T$.  Therefore this  model, hereafter referred to as the ``regular'' input model, represents spiking that is more regular than a Poisson process.

The synaptic response with the regular input model can be analyzed using methods similar to those used for the irregular model.  We introduce an auxiliary process, $q(t)$, that transitions sequentially through the state space $\{1,\ldots,\theta\}$.  Once reaching $\theta$, $q(t)$ transitions back to state 1.  Transitions occur as a Poisson process with rate $\theta r_{\textrm{in}}$.  The waiting times between transitions from $q=\theta$ to $q=1$  are gamma distributed.  Thus, to recover the regular input model, we specify that each  transition from $1=\theta$ to $q=1$ represents  a single presynaptic spike.  The process $g(t)=(m(t),q(t))$ is then a continuous time Markov chain on the discrete state space $\{1,\ldots,\theta\}\times \{0,\ldots,M\}$.  We enumerate all $\theta(M+1)$ elements of this space and denote the $j$th element as $\Gamma_j=(m_j,q_j)$ for $j=1,\ldots,\theta (M+1)$.

The infinitesimal generator, $G$, which is a $\theta (M+1)\times \theta (M+1)$ matrix is defined analogously to the matrix $B$ in Eq.~\eqref{E:B} above.  The elements of $G$ can be filled using the following transition probabilities.  
As for the irregular input model, vesicle recovery occurs as a Poisson process with rate $(M-m(t))/\tau_u$ so that 
$$
\lim_{h\to 0}\frac{1}{h}\Pr(g(t+h)=(m+1,q)\,|\,g(t)=(m,q))=\frac{M-m}{\tau_u}
$$
for $m=0,\ldots,M$ and $q=1,\ldots,\theta$.
Transitions that increment $q(t)$ occur with instantaneous rate, $\theta r_{\textrm{in}}$ so that
$$
\lim_{h\to 0}\frac{1}{h}\Pr(g(t+h)=(m,q+1)\,|\,g(t)=(m,q))=\theta r_{\textrm{in}}
$$
for $q=1,\ldots,\theta-1$ and $m=0,\ldots,M$.  The only other transitions are those from $q(t)=\theta$ to $q(t)=1$, which represent a presynaptic spike and are therefore accompanied by a release of vesicles.  The transitions contribute the following, 
\begin{align*}
&\lim_{h\to 0}\frac{1}{h}\Pr(g(t+h)=(1,m-k)\,|\,g(t)=(\theta,m))=\\
&\theta r_{\textrm{in}} \frac{m!}{(m-k)!}{p_r}^k(1-{p_r} )^{m-k}
\end{align*}
for $m\in\{0,1,\ldots,M\}$ and $k\in \{0,\ldots,m\}$.
These transition rates can be used to fill the off-diagonal terms of the matrix $G$.  The diagonal terms are then filled so that the rows sum to zero.
The stationary distribution, $p_0$, of $g(t)=(m(t),q(t))$ is given by the vector in the one-dimensional null space of $G$ with elements that sum to one.




A proportion $[p_0]_{\gamma(k)}$ of time is spent in state $(m(t),q(t))=(k,\theta)$ where $\gamma(k)$ represents the index of the element $(k,\theta)$ in the enumeration chosen for $\Gamma$ (i.e., the index, $j$, at which $\Gamma_j=(k,\theta)$).  In that state, the transition to $q(t)=1$ occurs with instantaneous rate $\theta r_{\textrm{in}}$ and releases average of ${p_r} m(t)$ vesicles.  Thus, the mean rate of vesicle release is given by  
$$
r_x=\sum_{k=1}^M  \theta r_{\textrm{in}} {p_r}  k \,[p_0]_{\gamma(k)}. 
$$

As above, we separate the auto-covariance into a delta function and a continuous part by writing
$
R_x(\tau)=A_x \delta(\tau)+R^+_x(\tau)
$
where $R^+_x(\tau)$ is a continuous function.  The area of the delta function at the origin is given by
$$
A_x=\sum_{k=1}^M \theta r_{\textrm{in}} [p_0]_{\gamma(k)} \left(k{p_r} (1-{p_r} )+k^2{p_r}^2\right)
$$
by an argument identical to that used for the irregular input model above.
Also by a similar argument used for the irregular input model, we have that
\begin{align*}
&R_x(\tau)=\\
&A_x\delta(\tau)-r_x^2+\theta^2r_{\textrm{in}}^2\sum_{k,l=1}^{M} [p_0]_{\gamma(k)}\left[e^{|\tau|G^T}\right]_{\gamma(l),\gamma(k)}  {p_r}^2 k l.
\end{align*}

\subsection{Parameters used in figures}

Theoretical results are obtained for arbitrary parameter values, but for all figures we use a set of parameter values that are consistent with experimental studies.  For synaptic parameters, we use $\tau_u=700$~ms and ${p_r} =0.5$ consistent with measurements of short term depression in pyramidal-to-pyramidal synapses in the rat neocortex~\cite{Tsodyks97,Fuhrmann02}.  We also choose $M=5$ which is within the range observed in several cortical areas~\cite{Branco09}.

The Poisson presynaptic input model is determined completely by its firing rate and the regular input model is determined completely by its firing rate and Fano factor.  Presynaptic firing rates and Fano factors are  reported on the axes or captions of each figure.  The irregular input model has four parameters that determine the firing rate and Fano factor.  In all figures, we set $\tau_b=\tau_s=1.315/c$, $r_b=37c$, and $r_s=3c$ which gives a Fano factor of $F_{\textrm{in}}=20.0017\approx 20$ for any value of $c$ (from Eq.~\eqref{E:FinBurst}).  Changing $c$ effectively scales the timescale of presynaptic spiking, hence scaling $r_{\textrm{in}}$, without changing $F_{\textrm{in}}$.






\section{Results}

We analyze the synaptic response to different patterns of presynaptic inputs using a stochastic model of short term synaptic depression in which a presynaptic neuron makes $M$ functional contacts onto a postsynaptic neuron~\cite{Vere1966,Fuhrmann02,Goldman04,Rocha05}.  The input to the presynaptic neuron is a spike train denoted by $I(t)$.  Neurotransmitter vesicles are released probabilistically in response to each presynaptic spike.  Specifically, a contact with a readily available vesicle releases this vesicle with probability ${p_r} $ in response to a single presynaptic spike.  After a synaptic contact has released its neurotransmitter vesicle, it enters a refractory state where it is unable to release again until the vesicle is replaced.  The duration of this refractory period is an exponentially distributed random variable with mean $\tau_u$, so that vesicle recovery is Poisson in nature.


We are interested in how the statistics of the presynaptic spike train determine the statistics of the synaptic response.  The presynaptic statistics are quantified using the presynaptic firing rate, $r_{\textrm{in}}$, the presynaptic auto-covariance function, $R_{\textrm{in}}(\tau)$, and the Fano factor, $F_{\textrm{in}}(T)$, of the number of presynaptic spikes during a window of length $T$.  Similarly, we quantify the statistics of the synaptic response using the mean rate of vesicle release, $r_x$, the auto-covariance of vesicle release, $R_x(\tau)$, and the Fano factor, $F_x(T)$, of the number of vesicles released during a window of length $T$.  We will especially focus on Fano factors over large time windows and define $F_{\textrm{in}}=\lim_{T\to\infty}F_{\textrm{in}}(T)$, $F_x=\lim_{T\to\infty}F_x(T)$ accordingly. See Methods for more details. 

We begin by considering the effect of $F_{\textrm{in}}$ on the mean rate of vesicle release, $r_x$.  We then examine the dependence of $F_x(T)$ on the length, $T$, of the time window over which vesicle release events are counted.  Finally, we show that short term synaptic depression promotes Poisson-like responses to  non-Poisson presynaptic inputs.

\subsection{Irregular presynaptic spiking reduces the rate at which neurotransmitter vesicles are released}

\begin{figure}
\includegraphics[width=84mm]{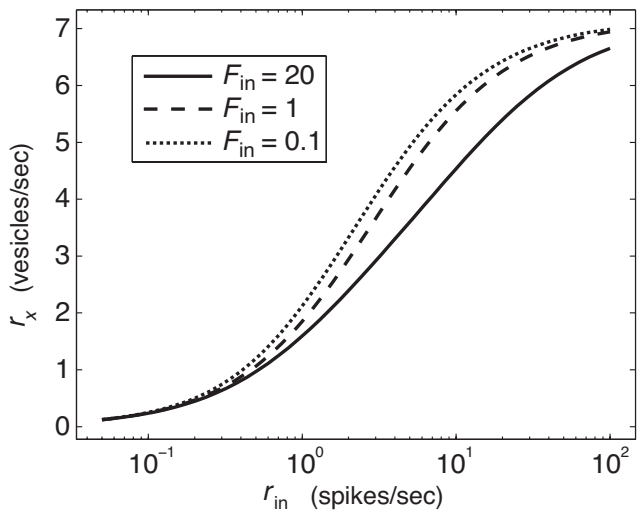}
\caption{{\bf Rate of vesicle release as a function of presynaptic firing rate for various presynaptic Fano factors.}  The rate of vesicle release, $r_x$, is  an increasing function of presynaptic firing rate, $r_{\textrm{in}}$.  Vesicle release is slower for the irregular spiking model than for the Poisson and regular spiking model.}
\label{F:rinvsrx}
\end{figure}

\begin{figure*}
\includegraphics[width=174mm]{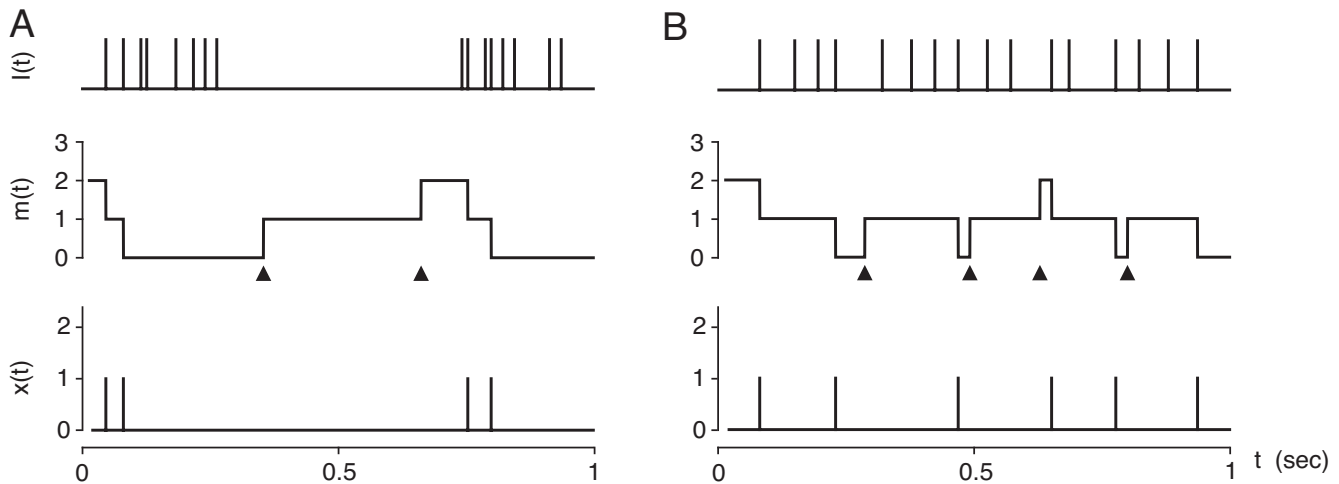}
\caption{{\bf Synaptic response to an irregular and a regular presynaptic spike train.}  {\bf A)} An irregular spike train, $I(t)$, drives a depressing synapse.  Each vesicle release event is indicated by a vertical bar with height indicating the number of vesicles released (here, all events release just one vesicle).  Each time a vesicle is released, the number of available vesicles, $m(t)$, is decremented accordingly.  Vesicle recovery increments $m(t)$ and occurs randomly in time (vesicle recovery events indicated by filled triangles).  {\bf B)} Same as (A) except for a more regular presynaptic spike train.  Note that, even though the same number of spikes occur in (A) and (B), the irregular spike train is less effective in releasing vesicles.  This occurs because all vesicles are depleted by the first few spikes in a burst and subsequent spikes in that burst are unable to release vesicles.  For illustrative purposes, we set $M=3$ in this figure.
}
\label{F:TracesBG}
\end{figure*}

We first briefly investigate the dependence of the rate of vesicle release, $r_x$, on the rate and   variability of the presynaptic spike train, as measured by $r_{\textrm{in}}$ and $F_{\textrm{in}}$ respectively.  Vesicle release rate generally increases with $r_{\textrm{in}}$, but saturates to $r_x=M/\tau_u$ whenever ${p_r} r_{\textrm{in}}\gg 1/\tau_u$ since synapses are depleted in this regime~(Fig.~\ref{F:rinvsrx}).

When presynaptic spike times occur as a Poisson process (so that $F_{\textrm{in}}=1$), the mean rate of vesicle release is given by $r_x=M{p_r} r_{\textrm{in}}/({p_r} r_{\textrm{in}} \tau_u+1)$~\cite{Fuhrmann02,Rocha05,RosenbaumPLoS12}.  Interestingly, vesicle release is slower for more irregular presynaptic spiking and faster for more regular presynaptic spiking even when presynaptic spikes arrive at the same mean rate (Fig.~\ref{F:rinvsrx}, also see \cite{Rocha02}).  This can be understood by noting that, for the irregular input model, spikes arrive in bursts of higher firing rate followed by durations of lower firing rate.  Vesicles are depleted by the first few spikes in a burst and subsequent spikes in that burst are ineffective and therefore essentially ``wasted'' spikes (Fig.~\ref{F:TracesBG}A).  When presynaptic spikes arrive more regularly, more vesicles are released on average~ (Fig.~\ref{F:TracesBG}B).

\subsection{Variability in the number of vesicles released in a time window decreases with window size}\label{S:FT}

We now consider how the the variability of the synaptic response to a presynaptic input depends on the timescale over which this variability is measured.  We quantify the variability of the synaptic response using the Fano factor, $F_x(T)$, which is defined to be the variance-to-mean ratio of the number of vesicles released in a time window of length $T$ (see Methods) and can be calculated from an integral of the auto-covariance function, $R_{x}(\tau)$, using Eq.~\eqref{E:VarIntx}.

\begin{figure*}
\includegraphics[width=174mm]{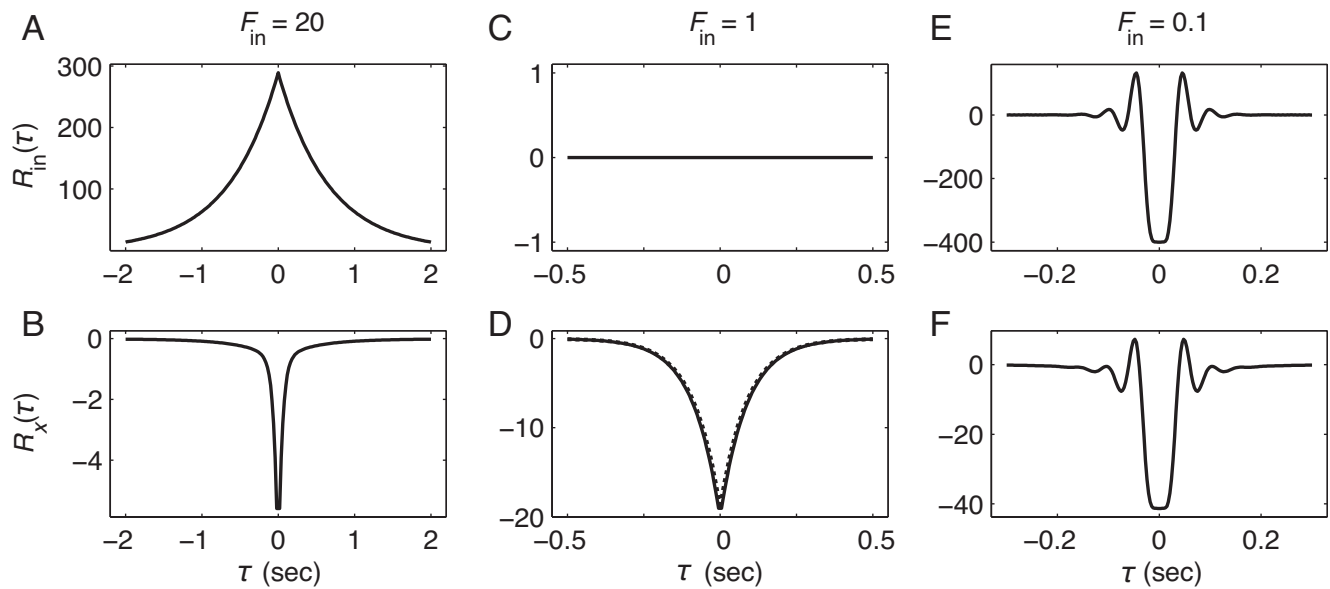}
\caption{{\bf Presynaptic auto-covariance functions and auto-covariance of vesicle release for three input models.}  Auto-covariance functions of presynaptic spike trains (top row) and the synaptic response they evoke (bottom row) for three different presynaptic input models: {\bf A,B)}irregular ($F_{\textrm{in}}=20$), {\bf C,D)} Poisson ($F_{\textrm{in}}=1$) and {\bf E,F)} regular ($F_{\textrm{in}}=0.1$).  Each auto-covariance function has a Dirac delta function at the origin that is not depicted here.   Dotted line in (D) is from the approximation in Eq.~\eqref{E:RxPoisson} and solid line is from exact calculation obtained using numerics for the regular input model with $\theta=1$ (see Methods), but the two are nearly indistinguishable.
$R_{\textrm{in}}(\tau)$ has units (spikes/sec)$^2$ and $R_x(\tau)$ has units (vesicles/sec)$^2$.  Short term depression introduces negative temporal correlations even when presynaptic spike trains are temporally uncorrelated (C,D) or positively correlated (A,B).}
\label{F:ACGs}
\end{figure*}

\begin{figure*}
\includegraphics[width=174mm]{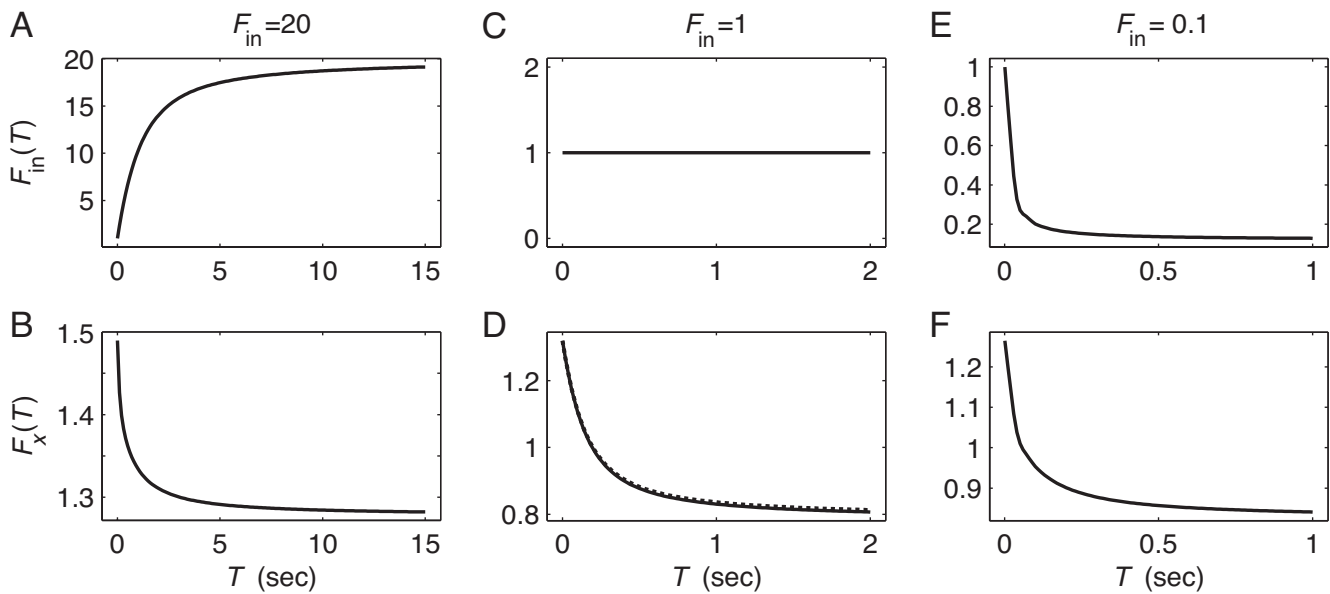}
\caption{{\bf Presynaptic and response Fano factor as a function of window size for three input models.}  Presynaptic and response Fano factors, $F_{\textrm{in}}(T)$ and $F_x(T)$, as a function of the window size over which inputs or vesicles are counted (see Methods), obtained by applying Eq.~\eqref{E:VarInt} to the auto-covariance functions in Fig.~\ref{F:ACGs}.  Short term depression causes response Fano factor to decrease with window size even when presynaptic Fano factor increases with window size (A,B) or is independent of window size (C,D).  Also, response Fano factors are near 1 even when presynaptic Fano factors are not (B and F). }
\label{F:FvsT}
\end{figure*}

The auto-covariance of a  Poisson presynaptic spike train is simply a delta function at the origin, $R_{\textrm{in}}(\tau)=r_{\textrm{in}} \delta(\tau)$, and the Fano factor over any window size is therefore equal to one, $F_{\textrm{in}}(T)=1$ (Figs. \ref{F:ACGs}C and \ref{F:FvsT}C).
The auto-covariance of the synaptic response when presynaptic inputs are Poisson consists of a delta function at the origin surrounded by a double-sided exponential with a negative peak (see Eq.~\eqref{E:RxPoisson} and Fig.~\ref{F:ACGs}D) that decays with timescale  $\tau_0={\tau_u}/{(1+{p_r} r_{\textrm{in}}\tau_u)}$.  
The fact that the auto-covariance is negative away from $\tau=0$ implies that the Fano factor, $F_x(T)$, is monotonically decreasing in the window size, $T$ (see Eq.~\eqref{E:FxPoisson} and Fig.~\ref{F:FvsT}D).
For small $T$, the mass of the delta function at the origin dominates the integral in Eq.~\eqref{E:VarIntx} so that the Fano factor is approximately equal to the ratio of this mass to the mean rate, $r_x$, at which vesicles are released.  As $T$ increases, the negative mass of the exponential peak subtracts from the positive contribution of the delta function and decreases the Fano factor.  In particular, $F_x(T)\approx D-ET+\mathcal O(T^2)$ where $Dr_x$ is the mass of the delta function in $R_x(\tau)$ and $-Er_x$ is the peak of the exponential in $R_x(\tau)$ (see Eqs.~\eqref{E:D} and \eqref{E:E}).  As $T$ continues to increase, $F_x(T)$ monotonically decreases towards its limit, $F_x:=\lim_{T\to\infty}F_x(T)=D-2E\tau_0$.  Thus, short term synaptic depression converts a Fano factor that is constant with respect to window size into one that decreases with window size (Fig.~\ref{F:FvsT}C,D).

When presynaptic spike times are not Poisson, the statistics of the postsynaptic response cannot be derived analytically using the methods utilized for the Poisson input model.  Instead, we use the fact that the synapse model can be represented using a continuous time Markov chain, which can be analyzed to derive expressions for the response statistics in terms of an infinitesimal generator matrix (see Methods).

Irregular presynaptic spiking (i.e., inputs with $F_{\textrm{in}}>1$) is achieved by varying the rate of presynaptic spiking randomly in time to produce a doubly stochastic Poisson process (see Methods). For this  model, the input auto-covariance is a delta function at the origin surrounded by an exponential peak (see Eq.~\ref{E:RinBurst} and Fig.~\ref{F:ACGs}A).  The input Fano factor therefore increases with window size~(see Eq.~\ref{E:FinBurst} and Fig.~\ref{F:FvsT}A).  The positive temporal correlations exhibited in the input auto-covariance function are canceled by the temporal de-correlating effects of short term synaptic depression~\cite{Goldman99,Goldman02,Goldman04}.  For the parameters chosen in this study, this de-correlation outweighs the positive presynaptic correlations so that the auto-covariance function of the response is negative away from $\tau=0$ (Fig.~\ref{F:ACGs}B), although parameters can also be chosen so that temporal correlations in the response are small and positive~\cite{Goldman02}.  As with the Poisson input model, negative temporal correlations cause the response Fano factor to decrease with window size~(Fig.~\ref{F:FvsT}B).  Thus short term synaptic depression and stochastic vesicle dynamics can convert a presynaptic Fano factor that \emph{increases} with window size into one that \emph{decreases}.

Regular presynaptic spiking is achieved by generating a renewal process with gamma-distributed interspike-intervals.  The input auto-covariance function for this model exhibits temporal oscillations (Eq.~\eqref{E:RinGamma} and Fig.~\ref{F:ACGs}E) and the Fano factor generally decreases with window size (Fig.~\ref{F:FvsT}E).  Perhaps unsurprisingly, the auto-covariance function of the synaptic response exhibits oscillations and the response Fano factor decreases with window size (Figs. \ref{F:ACGs}F and \ref{F:FvsT}F).

For all three input models, the variability of the synaptic response is larger over shorter time windows and smaller over larger time windows.  A postsynaptic neuron that is in an excitable regime will generally respond most effectively to inputs that exhibit more variability over short time windows~\cite{Salinas00,Salinas02,MorenoBote08,MorenoBote10}.  In addition, rate coding is often more efficient when spike counts over larger time windows are less variable~\cite{Zohary94}.  Thus, the dependence of $F_x(T)$ on window size is especially efficient for the neural transmission of rate-coded information~\cite{Goldman04}.

In addition to the temporal dependence of $F_x(T)$ introduced by short term depression, note that the response Fano factor for the irregular input model is substantially smaller than the input Fano factor (Fig.~\ref{F:FvsT}).  Conversely, the response Fano factor for the regular input model is larger than the input Fano factor (Fig.~\ref{F:FvsT}).  For both models, the response Fano factor is substantially nearer to 1 than the input Fano factor.  We explain this phenomenon next.

\subsection{Depleted synapses exhibit Poisson-like variability even when presynaptic inputs are highly non-Poisson}

\begin{figure}
\includegraphics[width=84mm]{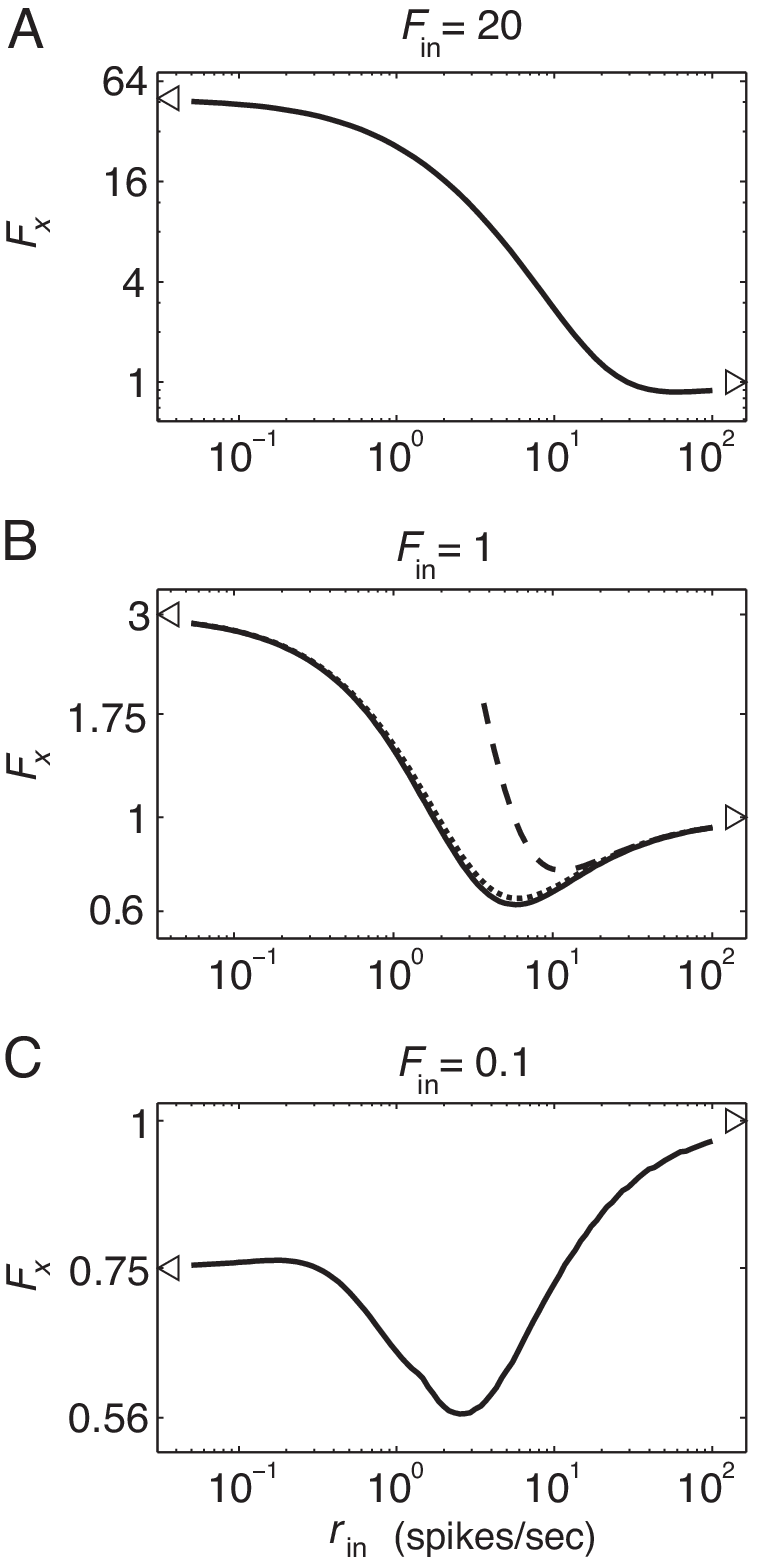}
\caption{{\bf Response Fano factor as a function of presynaptic firing rate for three input models.} Response Fano factors calculated over large windows for {\bf A)} the irregular input model {\bf B)} Poisson input model and {\bf C)} regular input model.  Fano factors
approach 1 at high presynaptic firing rates regardless of the presynaptic Fano factor (triangle on right is placed at $F_x=1$).  At low presynaptic firing rates, response Fano factors approach the value given in Eq.~\eqref{E:FxSlow} (indicated by triangle on left).  Dotted line in (B) is from closed form approximation in  Eq.~\eqref{E:FxInfPoisson} and dashed line is from the expansion given in Eq.~\eqref{F:Falpha}.
}
\label{F:rinvsFx}
\end{figure}

We now investigate the dependence of the variability in synaptic response on the rate and variability of the presynaptic input.  Since we have already discussed the dependence of $F_x(T)$ on $T$ above, we will focus here on the Fano factor calculated over long time windows, $F_x=\lim_{T\to\infty}F_x(T)$.

\begin{figure*}
\includegraphics[width=174mm]{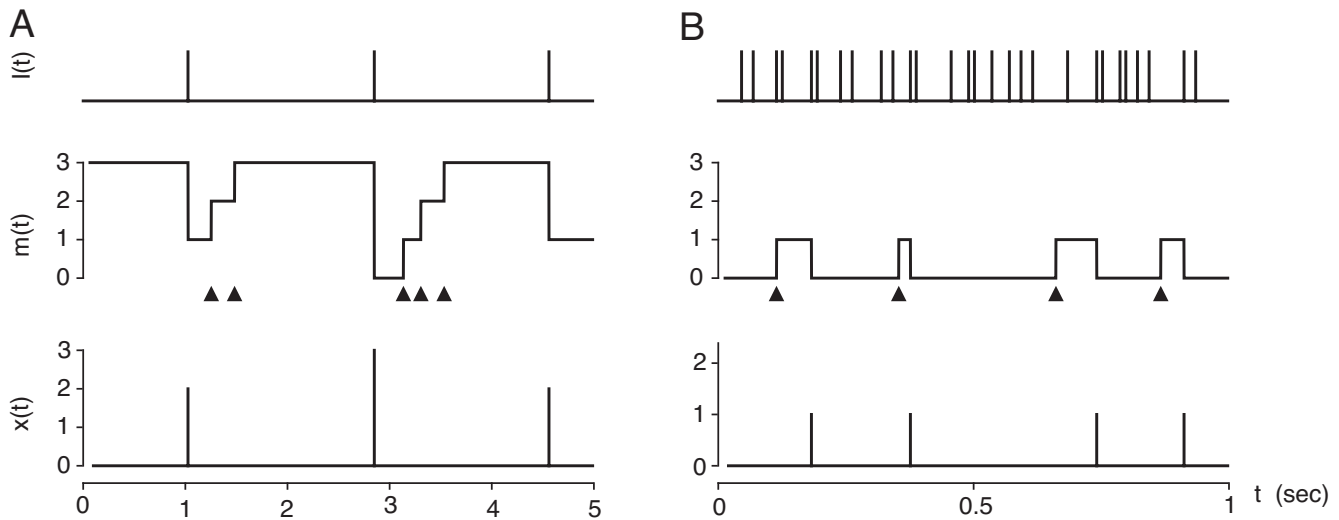}
\caption{{\bf Vesicle release dynamics at low and high presynaptic firing rates.}
{\bf A)} At low presynaptic rates, vesicles are recovered (i.e., $m(t)$ returns to $M=3$) between presynaptic spikes.  Thus, the number of vesicles released by each presynaptic spike is approximately an independent binomial random variable with mean ${p_r} M$ and variance ${p_r} M(1-M)$. {\bf B)} At high presynaptic rates, vesicles are released almost immediately after they are recovered.  Thus, the number of vesicles released over a time window of length $T$ is approximately a Poisson random variable with mean and variance $TM/\tau_u$.
}
\label{F:TracesSlowFast}
\end{figure*}

We first consider parameter regimes where the effective rate of presynaptic inputs is much slower than the rate of vesicle recovery (${p_r} r_{\textrm{in}}\ll 1/\tau_u$).  In such a regime, each contact is likely to recover between two consecutive presynaptic spikes and therefore all $M$ contacts are likely to have a vesicle ready to release when each spike arrives (Fig.~\ref{F:TracesSlowFast}A).  In this limit, the number of vesicles released by each spike is an independent binomial variable with mean $\langle w_j\rangle ={p_r} M$ and variance ${\textrm{var}}(w_j)=M{p_r} (1-{p_r} )$.  The number, $N_x(T)$, of vesicles released in a time window of length $T$ can then be represented as a sum of $N_{\textrm{in}}(T)$ independent binomial random variables (i.e., a random sum).  The mean of this sum is given by $\langle N_x(T)\rangle=\langle N_{\textrm{in}}(T)\rangle \langle w_j\rangle$, which implies that $r_x=M{p_r} r_{\textrm{in}}$ in this limit.  Similarly, the variance of this sum is given by~\cite{Karlin1} ${\textrm{var}}(N_x(T))=\langle N_{\textrm{in}}(T)\rangle {\textrm{var}}(w_j)+\langle w_j\rangle^2{\textrm{var}}(N_{\textrm{in}}(T))$, which implies
\begin{align}\label{E:FxSlow}
\lim_{r_{\textrm{in}}\to 0}F_x(T)&=\langle w_j\rangle F_{\textrm{in}}(T)+{\textrm{var}}(w_j)/\langle w_j\rangle\\
&=1+{p_r} (MF_{\textrm{in}}(T)-1).\notag
\end{align}
Eq.~\eqref{E:FxSlow} is verified for the Poisson input model by taking $r_{\textrm{in}}\to 0$ in Eqs.~\eqref{E:FxPoisson}.
For the irregular and regular input models, Eq.~\eqref{E:FxSlow} should be interpreted heuristically, as it was derived heuristically.  A counterexample to Eq.~\eqref{E:FxSlow} for the irregular input model can be constructed by fixing $r_f$ and $\tau_f$, then letting $\tau_s\to\infty$ and $r_s\to 0$ to achieve the $r_{\textrm{in}}\to 0$ limit.  In this case, our assumption that each contact is increasingly likely to recover between two consecutive spikes is violated and Eq.~\eqref{E:FxSlow} is not valid (not pictured).  Regardless, we verify numerically that Eq.~\eqref{E:FxSlow} is accurate when $r_{\textrm{in}}$ is decreased toward zero while keeping $F_{\textrm{in}}$ fixed (Fig.~\ref{F:rinvsFx}).

We now discuss the statistics of the postsynaptic response  when the effective presynaptic spiking is much faster than vesicle recovery (${p_r} r_{\textrm{in}}\gg 1/\tau_u$).  
In such a regime, incoming spikes occur much more frequently than recovery events and  synapses becomes depleted.
As a result, the number of vesicles released over a long time window is determined predominantly by the number of recovery events in that time window and largely independent from the number of presynaptic spikes (Fig.~\ref{F:TracesSlowFast}B)  \cite{Rocha05,RosenbaumPLoS12}.  The synaptic response therefore inherits the Poisson statistics of the recovery events so that $$\lim_{r_{\textrm{in}}\to\infty}F_x(T)=1.$$  For the Poisson input model, this limit can be made more precise in the $T\to\infty$ limit by expanding Eq.~\eqref{E:FxInfPoisson} in terms of the parameter $\alpha=1/({p_r} r_{\textrm{in}}\tau_u)$ to obtain
\begin{equation}\label{F:Falpha}
F_x=1-2\alpha+4\alpha^2+\mathcal O(\alpha^3)
\end{equation}
which converges to 1 as $r_{\textrm{in}} \tau_u\to\infty$.  For the irregular and regular input models, we verify in Fig.~\ref{F:rinvsFx} that $F_x\to 1$ when $r_{\textrm{in}}$ is increased while keeping $F_{\textrm{in}}$ fixed.

The time constant, $\tau_u$, at which a synapse recovers from short term depression has been measured in a number of experimental studies and is often found to be several hundred milliseconds~\cite{Tsodyks97,Varela97,Markram98,Galarreta98,Fuhrmann02,Hanson02,RavAcha05}.  Therefore, for even moderate presynaptic firing rates, synapses are often in a highly depleted state.  As discussed above, this promotes Poisson-like variability in the synaptic response.  This provides one possible mechanism through which irregular Poisson-like firing can be sustained in neuronal populations~\cite{Churchland10}.






\section{Discussion}

We used continuous time Markov chain methods to derive the response statistics of a stochastic model of short term synaptic depression with three different presynaptic input models.  We then used this analysis to understand how the mean presynaptic firing rate and the variability of presynaptic spiking interact with  synaptic dynamics to determine the mean rate of vesicle release and  variability in the number of vesicles released.  This analysis revealed a number of fundamental, qualitative dependencies of response statistics on presynaptic spiking statistics.  Some of the dependencies have been previously noted in the literature and some have not.

The number of vesicles released over a time window is smaller for irregular inputs than for more regular inputs (Figs. \ref{F:rinvsrx} and \ref{F:TracesBG}) given the same number of presynaptic spikes.  Thus, regular presynaptic spiking is more efficient at driving synapses.  This mechanism competes with a well-known property of excitable cells: that they are driven more effectively by irregular, positively correlated synaptic input currents~\cite{Salinas00,Salinas02,MorenoBote08,MorenoBote10}.  In addition, a \emph{population} of presynaptic spike trains drives a postsynaptic neuron more efficiently when the population-level activity is more irregular, for example due to pairwise correlations~\cite{Rocha05}.  Together, these results suggest that a postsynaptic neuron is most efficiently driven by presynaptic populations that exhibit small or negative auto-correlations, but positive pairwise cross-correlations.

Our model predicts that the de-correlating effects of short term depression and stochastic vesicle dynamics can produce negative temporal auto-correlations in the synaptic response even when presynaptic spiking is temporally uncorrelated or positively correlated, in agreement with previous studies~\cite{Goldman02,Rocha05}.  This yields a response Fano factor that decreases with window size, as observed in some recorded data~\cite{Kara00}.  We note, though, that some parameter choices can yield positively a correlated synaptic response when presynaptic  inputs are positively correlated~\cite{Goldman02} and neuronal membrane dynamics can introduce positive correlations to a postsynaptic spiking response even when synaptic currents are not positively correlated in time~\cite{MorenoBote06}.  This is consistent with several studies showing positive temporal correlations in recorded spike trains~\cite{Bair94,Dan96,Baddeley97,Churchland10}.

We predict that moderate or high firing rates can induce a Poisson-like synaptic response even when presynaptic inputs are non-Poisson (Fig.~\ref{F:rinvsFx} and \cite{Rocha02}).  This is because even moderate firing rates can deplete synapses and depleted synapses inherit the Poisson-like variability of synaptic vesicle recovery (Fig.~\ref{F:TracesSlowFast}B and \cite{Rocha05,RosenbaumPLoS12}).   At lower firing rates, short term depression and synaptic variability can increase or decrease Fano factor.  For example, in Fig.~\ref{F:rinvsFx}B, the response Fano factor is larger than the presynaptic Fano factor ($F_{\textrm{in}}=1$) at low firing rates, decreases at higher firing rates, then approaches $F_x=1$ at higher firing rates.  This complex dependence of firing rate on Fano factor might be related to the stimulus dependence of Fano factors observed in several cortical brain regions~\cite{Churchland10}.

Our conclusion that fast presynaptic spiking causes Poisson-like variability in the synaptic response relied on the assumption that vesicle recovery times are exponentially distributed.  The exponential distribution is a justifiable choice for recovery times only if recovery times obey a memoryless property: having already waited $t$ units of time for a recovery event, the probability of waiting an additional $s$ units of time does not depend on $t$.  The precise mechanics of vesicle re-uptake and docking determine whether this is an appropriate assumption.  If recovery times have a different probability distribution, then the synaptic response will inherit the properties of this distribution at high presynaptic firing rates instead of inheriting the Poisson-like nature of exponentially distributed recovery times.

Previous methods have been developed to analyze the synaptic response of the model used here.  In~ \cite{Rocha02,Goldman04},  the model restricted to the $M=1$ case is analyzed for presynaptic spike trains that are renewal processes.  This includes the Poisson and the regular input model discussed here, but excludes the irregular input model in which the spike train is a non-renewal inhomogeneous Poisson process.  In~\cite{RosenbaumPLoS12} approximations are obtained for the case where the presynaptic spike train is an inhomogeneous Poisson process, but the approximation is only valid when the rate-modulation of the Poisson process is small compared to the average firing rate.  Thus, these approximations are only valid for the irregular input model when $r_f-r_s\ll r_s$.  Other studies~\cite{Lindner09,Merkel10} use a deterministic synapse model that implicitly treats the number of available vesicles as a continuous rather than a discrete quantity.  This deterministic model represents the trial average of the model considered here and can vastly underestimate the variability of a synaptic response~\cite{RosenbaumPLoS12}.

A more detailed synapse model  allows for multiple  docking sites at a single contact~\cite{Wang99,Rocha05}.  This model can yield different response properties than the model used here in certain parameter regimes~\cite{Rocha05}.  Even though this more detailed model can be represented as a continuous-time Markov chain, the analysis of this model would be significantly more complex than the analysis considered here since it would be necessary to keep track of the number of readily releasable vesicles at each contact separately.  This would result in a Markov chain with $K\times N^M$ states where $M$ is the number of contacts, $N$ is the number of docking sites per contact and $K$ is the number of states used for the presynaptic input model ($K=1$ for the Poisson input model, $K=2$ for the irregular input model, and $K=\theta$ for the regular input model).

To quantify the synaptic response to a presynaptic spike train, we focused on the statistics of the number of vesicles released in a time window.  Postsynaptic neurons observe changes in  synaptic conductance in response to presynaptic spikes.  The synaptic conductance are often modeled in such a way that they can be easily derived from our process $x(t)$ through a convolution: $g(t)=\int_0^t x(t-s)\alpha(s)ds$ where $g(t)$ is the synaptic conductance elicited by a presynaptic spike train and $\alpha(s)$ is a kernel representing the characteristic postsynaptic conductance elicited by the release of a single neurotransmitter vesicle.  Since this mapping is linear, the statistics of $g(t)$ can easily be derived in terms of the statistics of $x(t)$~\cite{Tetzlaff08,RosenbaumPLoS12}.

%
%


\bibliographystyle{plain}      

\end{document}